\documentclass{nsart_eng}
\usepackage{cite}
\usepackage[dvips]{graphicx}
\usepackage{amsmath,amssymb}
\usepackage{enumitem}

\year{2023} \volume{0} \nomer{0} \firstpage{1}

\let\le\leqslant

\let\geq\geqslant

\usepackage{graphicx}
\usepackage{xcolor}
\usepackage{subcaption}
\newcommand{\be}{\begin{equation}}
\newcommand{\ee}{\end{equation}}
\newcommand{\ba}{\begin{array}{l}}
\newcommand{\ea}{\end{array}}
\newcommand{\re}[1]{(\ref{#1})}
\newcommand{\ci}[1]{\cite{#1}}
\newcommand{\banonum}{\begin{eqnarray*}}
\newcommand{\eanonum}{\end{eqnarray*}}
\newcommand{\baa}{\begin{eqnarray}}
\newcommand{\eaa}{\end{eqnarray}}
\newcommand{\bfr}{\begin{flushright}}
\newcommand{\efr}{\end{flushright}}
\newcommand{\bfl}{\begin{flushleft}}
\newcommand{\efl}{\end{flushleft}}
\newcommand{\lab}[1]{\label{#1}}

\begin{document}
\title[Fast forward evolution in heat equation]{Fast forward evolution in heat equation:\\ Tunable heat transport in adiabatic regime}
\author[J.~Matrasulov, J.R.~Yusupov, A.A.~Saidov]{$^{1}$J.~Matrasulov, $^{1}$J.R.~Yusupov and $^{2}$A.A.~Saidov}
%\affiliation{${^a}$ Kimyo International University in Tashkent, 156 Usman Nasyr Str., 100121, Tashkent, Uzbekistan\\
%${^b}$Turin Polytechnic University in Tashkent, 17 Niyazov Str.,100095,  Tashkent, Uzbekistan}
\address{
$^1$ Kimyo International University in Tashkent,\\ 156 Usman Nasyr Str., 100121, Tashkent, Uzbekistan\\
$^2$ Turin Polytechnic University in Tashkent,\\ 17 Niyazov Str.,100095,  Tashkent, Uzbekistan}

\email{jasur1362@gmail.com}

 % insert PACS

\begin{abstract}
 We consider the problem of fast forward evolution of the processes described in terms of the heat equation. The matter is considered on an adiabatically expanding time-dependent box. Attention is paid to acceleration of heat transfer processes. So called  shortcuts to adiabaticity, implying fast forwarding of the adiabatic states are studied. Heat flux and temperature profiles are analyzed for standard and fast forwarded regimes.    
\end{abstract}
\maketitle

\section{Introduction}
Controlling the evolution of physical processes is of fundamental and practical importance for modern science and technology. Mathematically, the problem of tunable evolution of a given physical system can be solved by developing realistic and physically acceptable models allowing manipulations in the evolution equations describing the process. Such a task attracted special attention in quantum mechanics, where an effective prescription for acceleration (deceleration) of a quantum evolution in the Schr\"odinger equation  has been proposed earlier in \ci{mas1}.
The advantage of the prescription proposed in \ci{mas1} is caused by the fact that it allows to speed up the time evolution of a wave function by controlling the driving potential with resultant regulation of the additional phase of the wave-function. Slightly modified version of the prescription was applied also to acceleration of the evolution for nonlinear Schr\"odinger equation and quantum tunneling dynamics \cite{rv2-kn}.
Later the method was considerably improved \ci{mas2,MN11} and applied for different systems (see, e.g., Refs.\ci{rv2-kn} -\ci{Thermo}). Another important aspect of fast-forward protocol proposed in \ci{mas1} is the flexibility for the choice of control parameter that allows to apply it to the broad variety of physical systems described in terms of different evolution equations. In particular, the prescription was quite effective for fast forwarding of the adiabatic evolution \ci{mas2, MN11}. In case of the adiabatic evolution, the central problem is fast forward of the adiabatic dynamics within a given short time-scale, so called ``short-cuts'' to adiabaticity.
Simply, shortcuts to adiabaticity (STA) are the fast control protocols to drive the dynamics of system, which are fast routes to the final results of slow, adiabatic changes of the controlling parameters of a system \ci{Muga,Muga1}. In other words, a technique which allows to control dynamics of the adiabatic processes is called ``shortcuts to adiabaticity''.  The concept of STA was introduced first in the Ref.~\ci{Chen}. So far, several types of STA control protocols, such as counterdiabatic driving \ci{Rice}, (also known as transitionless quantum driving \ci{Berry}), inverse engineering \ci{Inver}, local counterdiabatic driving \ci{Campo} and fast-forward \ci{mas2,MN11} protocols have been developed. Some other approaches to the STA beyond the protocol proposed in \ci{mas1} are discussed in detail in the Ref.~\cite{Muga}. In this paper we use fast-forward approach STA from the Refs.~\ci{mas2,MN11} to develop STA protocol for the heat transport processes described in terms of the heat equation. It is important to note that the protocol was found as successful application to the non-equilibrium equation of states for the quantum gas under a rapidly moving  piston \cite{JM}. Also, it lead to a simple protocol to accelerate the adiabatic quantum dynamics of spin clusters \cite{iwan} and adiabatic control of tunneling states \cite{naka}. The fast-forward theory is also applicable to dynamical construction of classical adiabatic invariant \cite{jar} and to classical stochastic Carnot-like heat engine \cite{Thermo}. Here we will study temperature profile and heat current for fast forward system described in terms of the heat equation on a finite interval with moving boundaries.  This paper is organized as follows. In the next section we will present brief description of the fast-forward protocol, following the Ref \ci{mas1}. Section III presents brief recall of the heat equation on a finite interval. In section IV, application of fast forward protocol to the heat equation in a time-dependent box is described. Finally, the section V presents some concluding remarks. 

\section{Fast forward of adiabatic dynamics described in terms of parabolic evolution equations}\label{sec2}

Here, following the Ref.~\ci{mas1} we will briefly recall description of the fast forward protocol for the Schr\"odinger equation, which is a parabolic equation.  We do this for 1D confined system by focusing on adiabatic dynamics. First, let us  define two systems, the so-called standard system, $\Psi_{0}$, whose evolution  is to be fast accelerated and the fast forwarded system, $\Psi_{FF}$ which is just obtained as a result of acceleration of the eolution of the standard system. Our approach can be formulated as follows:
\begin{enumerate}[label=(\roman*)]
\item A given confining potential $V_{0}$ is assumed to vary in time adiabatically and to generate a stationary system $\psi_{0}$, which is an eigenstate of the time-independent Schr\"{o}dinger equation with the instantaneous Hamiltonian. Then both $\psi_{0}$ and $V_{0}$ are regularized so that they should satisfy the time-dependent Schr\"{o}dinger equation (TDSE);
\item Taking the regularized system as a standard system (to be fast forwarded), we shall change the time scaling with use of the scaling factor $\alpha\left(t\right)$, where the mean value $\bar{\alpha}$ of the infinitely-large time scaling factor, $\alpha(t)$ will be chosen to compensate the infinitesimally-small growth rate, $\epsilon$ of the quasi-adiabatic parameter and to satisfy $\bar{\alpha} \times \epsilon=\text{finite}$.
\end{enumerate}

Thus, consider the standard dynamics with a ``deformable'' trapping potential, whose shape is characterized by a slowly-varying control parameter $R(t)$ given by
\begin{eqnarray}
\label{2.1}\label{adia-R}
R(t)=R_{0}+\epsilon t,
\end{eqnarray}
with the growth rate $\epsilon\ll1$, which means that it requires a very long time $T=O\left(\frac{1}{\epsilon}\right)$, to see the recognizable change of $R(t)$. The dynamics of a charged particle confined in the field of such potential can be described in terms of the following time-dependent 1D Schr\"{o}dinger equation (1D TDSE):
\begin{eqnarray}
\label{2.2}
\textit{i}\hbar\frac{\partial\psi_{0}}{\partial t}=-\frac{\hbar^{2}}{2m}\partial_x^{2}\psi_{0}+V_{0}(x,R(t))\psi_{0},
\end{eqnarray}
where the coupling with external electromagnetic field is assumed to be absent. The stationary bound state $\phi_{0}$ satisfies 
the time-independent counterpart given by
\begin{eqnarray}
\label{2.3}
E\phi_{0}=\hat{H}_0\phi_{0}\equiv\left[-\frac{\hbar^{2}}{2m}\partial_x^{2}+V_{0}(x,R)\right]\phi_{0}.
\end{eqnarray}

Then, using the eigenstate $\phi_0=\phi_0(x, R)$, satisfying Eq.(\ref{2.3}),
one might conceive the corresponding time-dependent system to be a product of $\phi_0$ and a dynamical factor as,
\begin{eqnarray}\label{eq3.4}
\psi_{0}=\phi_{0}(x,R(t))e^{-\frac{\textit{i}}{\hbar}\int^{t}_{0}E(R(t'))dt'}.
\end{eqnarray}
As it stands, however, $\psi_{0}$ does not satisfy TDSE (\ref{2.2}). Therefore we introduce a regularized system
\begin{eqnarray}\label{eq3.5}
\psi^{reg}_{0}
&\equiv& \phi_{0}(x,R(t))e^{\textit{i}\epsilon\theta(x,R(t))}e^{-\frac{\textit{i}}{\hbar}\int^{t}_{0}E(R(t'))dt'}
\nonumber\\
&\equiv& \phi_0^{reg} (x,R(t))e^{-\frac{\textit{i}}{\hbar}\int^{t}_{0}E(R(t'))dt'}
\end{eqnarray}
together with a regularized potential
\begin{eqnarray}\label{eq3.6}
V^{reg}_{0}\equiv V_{0}(x,R(t))+\epsilon\tilde{V}(x,R(t)).
\end{eqnarray}
The unknown $\theta$ and $\tilde{V}$ will be determined self-consistently, so that $\psi^{reg}_{0}$ should fulfill the TDSE given by
\begin{eqnarray}\label{eq3.7}
\textit{i}\hbar\frac{\partial\psi_{0}^{reg}}{\partial t}=-\frac{\hbar^{2}}{2m}\partial_x^{2}\psi^{reg}_{0}+V^{reg}_{0}\psi^{reg}_{0},
\end{eqnarray}
up to the order of $\epsilon$.

One can rewrite $\phi_{0}(x,R(t))$ using the real positive amplitude $\overline{\phi}_{0}(x,R(t))$ and phase $\eta(x,R(t))$ as
\begin{eqnarray}\label{2.8}
\phi_{0}(x,R(t))=\bar{\phi}_{0}(x,R(t))e^{\textit{i}\eta(x,R(t))},
\end{eqnarray}
and see that $\theta$ and $\widetilde{V}$ satisfy the following relations:
\begin{eqnarray}\label{2.9}
%\partial_x^{2}\theta&+&2\partial_x(\ln\bar{\phi}_{0})\cdot\partial_x\theta \nonumber\\
\partial_x(\bar{\phi}_{0}^2\partial_x\theta) =-\frac{m}{\hbar}\partial_{R}\bar{\phi}_{0}^2,
\end{eqnarray}
\begin{eqnarray}\label{2.10}
\frac{\tilde{V}}{\hbar}=-\partial_{R}\eta-\frac{\hbar}{m}\partial_x\eta\cdot\partial_x\theta.
\end{eqnarray}
Integrating Eq.~(\ref{2.9}) over $x$, we have
\begin{equation}\label{2.11}
\partial_x\theta=-\frac{m}{\hbar}\frac{1}{\bar{\phi}_0^2}\int^x \partial_R\bar{\phi}_0^2 dx', 
\end{equation}
which is the main equation of the regularization procedure. The problem of singularity due to nodes of $\bar{\phi}_{0}$ in Eq.~(\ref{2.11}) can be overcome, so long as one is concerned with the systems with scale-invariant potentials.

We now accelerate the quasi-adiabatic dynamics of $\psi_{0}^{reg}$ in Eq.(\ref{eq3.5}), by applying the external electromagnetic field. To do this, we introduce the fast-forward version 
of $\psi_{0}^{reg}$ as
\begin{eqnarray}\label{2.12}
\psi_{FF}^{(0)}(x,t)&\equiv& \psi^{reg}_{0}(x,R(\Lambda(t)))\nonumber\\
&\equiv& \phi^{reg}_{0}(x,R(\Lambda(t)))e^{-\frac{\textit{i}}{\hbar}\int^{t}_{0}E(R(\Lambda(t')))dt'}\nonumber\\
\end{eqnarray}
with
\begin{equation}\label{rv2-1}
R(\Lambda(t))=R_0+\epsilon \Lambda(t),
\end{equation}
where $\Lambda(t)$ is the future or advanced time 
\begin{eqnarray}\label{1.3}
\Lambda(t)=\int_0^t\mathrm{\alpha(t')}\,\mathrm{d}t',
\end{eqnarray}
and $\alpha (t)$ is a magnification scale factor defined by $\alpha(0)=1$, $\alpha(t)>1$  $(0<t<T_{FF})$, $\alpha(t)=1$ $(t\geq T_{FF})$.
Suppose, $T$ to be a very long time to see a recognizable change of the adiabatic parameter $R(t)$ in Eq.(\ref{adia-R}), and then the corresponding change of $R(\Lambda(t))$ is reached in the shortened or fast-forward time $T_{FF}$ defined by
\begin{eqnarray}\label{T-Tff}
T=\int_0^{T_{FF}}\alpha (t) \mathrm{d}t.
\end{eqnarray}

The explicit expression for $\alpha(t)$ in the fast-forward range ($0 \le t \le T_{FF}$) can be written as
\begin{eqnarray} 
\label{1.7}
\alpha(t)=\bar{\alpha}- (\bar{\alpha}-1)\cos (\frac{2\pi}{T/\bar{\alpha}}t),
\end{eqnarray}
where $\bar{\alpha}$ is the mean value of $\alpha(t)$ and is given by $\bar{\alpha}=T/T_{FF}$.

Furthermore, let us assume $\psi_{FF}^{(0)}$ to be the solution of the TDSE for a charged particle in the presence of gauge potentials, 
$A_{FF}^{(0)}(x,t)$ and $V_{FF}^{(0)}(x,t)$,
\begin{eqnarray}
\label{1.9}
&&\imath\hbar\frac{\partial{\psi_{FF}^{(0)}}}{\partial{t}}=H_{FF}\psi_{FF}^{(0)}\equiv \nonumber\\
&&\left(\frac{1}{2m}(\frac{\hbar}{i}\partial_x - A_{FF}^{(0)})^2+ V_{FF}^{(0)}+V_0^{reg}\right)\psi_{FF}^{(0)},\nonumber\\
\end{eqnarray}
where, for simplicity we employ the prescription of a positive unit charge ($q=1$) and the unit velocity of light ($c=1$). The driving electric field is given by,
\begin{eqnarray}
\label{1.10}
E_{FF}=-\frac{\partial A_{FF}^{(0)}}{\partial t}-\partial_x V_{FF}^{(0)}.
\end{eqnarray}

Substituting Eq.~(\ref{2.12}) into Eq.~(\ref{1.9}), we find $\phi^{reg}_{0}$ to satisfy
\begin{eqnarray}
\label{2.13}
\textit{i}\hbar\frac{\partial\phi^{reg}_{0}}{\partial t}&=&\frac{1}{2m}\left(\frac{\hbar}{\textit{i}}\partial_x-A_{FF}^{(0)}\right)^{2}\phi^{reg}_{0} \nonumber\\
&+&(V_{FF}^{(0)}+V_0^{reg}-E)\phi^{reg}_{0},
\end{eqnarray}
where $V_0^{reg}\equiv V^{reg}(x,R(\Lambda(t)))$, i.e., the advanced-time variant of Eq.~(\ref{eq3.6}).
The dynamical phase in Eq.(\ref{2.12}) has led to the energy shift in the potential in Eq.~(\ref{2.13}).

Rewriting $\phi^{reg}_{0}$ in terms of the amplitude $\bar{\phi}_{0}$ and phases $\eta+\epsilon\theta$ as
\begin{eqnarray}\label{eq3.13}
\phi^{reg}_{0}\equiv  \bar{\phi}_{0}(x,R(\Lambda(t)))e^{\textit{i}\left[\eta(x,R(\Lambda(t)))
+\epsilon\theta(x,R(\Lambda(t)))\right]}, \nonumber\\
\end{eqnarray}
and using Eq.~(\ref{eq3.13}) in Eq.~(\ref{2.13}), we find $A_{FF}^{(0)}$ of $O(\epsilon\alpha)$ and $V_{FF}^{(0)}$ consisting of terms of $O(\epsilon\alpha)$ and $O((\epsilon\alpha)^2)$.     

Now, taking the limit $\epsilon\rightarrow 0$ and $\bar{\alpha}\rightarrow \infty$ with $\epsilon\bar{\alpha}=\bar{v}$ being kept finite, we obtain (see, the Ref.~\cite{naka} for details):
\begin{eqnarray}
\label{2.20}
A_{FF}^{(0)}&=&-\hbar v(t)\partial_x\theta,\nonumber\\
V_{FF}^{(0)}&=&-\frac{\hbar^{2}}{m}v(t)\partial_x\theta\cdot\partial_x\eta \nonumber\\
&-&\frac{\hbar^{2}}{2m}(v(t))^{2}(\partial_x\theta)^{2}-\hbar v(t)\partial_{R}\eta,
\end{eqnarray}
where  $T_{FF}\left(=\frac{T}{\bar{\alpha}}=O\left(\frac{1}{\epsilon\bar{\alpha}}\right)\right)=\text{finite}$.
In the same limiting case as above, $\psi_{FF}^{(0)}$ is explicitly given by
\begin{eqnarray}
\label{2.23}
\psi_{FF}^{(0)}=\bar{\phi}_{0}(x,R(\Lambda(t)))
e^{\textit{i}\eta(x,R(\Lambda(t)))}e^{-\frac{\textit{i}}{\hbar}\int^{t}_{0}E(R(\Lambda(t')))dt'}. \nonumber\\
\end{eqnarray}

\section{Fast forwarding the heat equation based evolution}

Our aim is application of fast forward protocol to the heat equation in time-varying interval, to achieve acceleration (deceleration) of the heat transfer process. Before formulating the problem, we will briefly recall heat equation and its solution considering it on a finite interval, $(0,l)$. The heat equation can be written as
\begin{equation}
    \frac{\partial u}{\partial t}=\kappa^2\frac{\partial^2 u}{\partial x^2}
\end{equation}
where $u(x,t)$ is the temperature profile, $\kappa$ is the heat conductivity. The boundary and initial conditions can be imposed as
\begin{equation}
    u(0,t)=u(l,t)=0,
\end{equation}
and
\begin{equation}
    u(x,0)=f(x).
\end{equation}
Exact solutions of the problem given by Eqs.~(1)-(3) can be obtained by factorizing variables as
\begin{equation}
    u(x,t)=X(x)T(t)=C_n\sin{\frac{\pi n}{l}x},
\end{equation}
Applying the principle of superposition gives the general solution of Eq.(23) \ci{HEBook1,HEBook2}
\begin{equation}
    u(x,t)=\sum_{n=1}^{\infty}C_n e^{-\frac{\pi^2\kappa^2n^2}{l^2}t}\sin{\frac{\pi n}{l}x}
\end{equation}
where $C_n$ can be calculated by using initial temperature profile as follows \ci{HEBook1,HEBook2}:
$$
C_n=\frac{2}{l}\int\limits_{0}^{l}f(x)\sin{\frac{\pi n}{l}x}dx.
$$
In what follows we will choose Gaussian type initial temperature profile given by
$$
f(x)=\frac{1}{\sqrt{2\pi}\sigma}e^{-\frac{(x-x_0)^2}{\sigma^2}}
$$
For such initial temperature profile, the coefficients, $C_n$ are computed numerically both for standard as well as for fast-forwarded systems.

Now consider the fast forward evolution in the heat equation. The standard system, $u^{(0)}$ is given in terms of the  following evolution equation:
\begin{equation}
\frac{\partial u^{(0)}}{\partial t}=\kappa^2\frac{\partial^2 u^{(0)}}{\partial x^2}
\end{equation}
where $\kappa$ denotes the thermal conductivity. 
We consider heat equation on a finite-interval with time-varying boundary. The time-dependent boundary conditions are imposed as
$$
u^{(0)}(x=0,t)=0, \;\; u^{(0)}(x=L(\Lambda(t)),t)=0,
$$
We consider the case of time-dependence of the boundary given by
$$
L(t)=L_0+\epsilon t.
$$
Furthermore, we assume that our physical process is adiabatic, i.e. $\epsilon\to 0$.
Then $t$ can be replaced by future time given as $\Lambda(t)=\int\limits_{0}^{t}\alpha(\tau)d\tau$. This allows us to write $L(t)$ as
$$
L(t)=L_0+\epsilon \Lambda(t).
$$
It is easy to see that wall's position for fast-forwarded system can have arbitrary time-dependence, since the magnification factor $\alpha$ can be any function of time. The velocity of the wall, whose value is finite, is given by
$$
\dot L=\epsilon \alpha=v, \; \; \epsilon \to 0, \; \; \alpha \to \infty.
$$
General solution of heat equation for adiabatically expanding box can be written as

\begin{equation}
u^{(0)}(x)=\sum_{n=1}^\infty C_n\sin{\Biggl[\frac{\pi n}{L(\Lambda(t))}x\Biggr]},
\lab{sol01}
\end{equation}
Before applying fast forward protocol of the Ref.~\ci{mas1}, we will regularize the standard system as it was done in \cite{MN11}. Then the solution of the regularized standard system can be written as
\begin{equation}
u_n^{(reg)}=u_n^{(0)}(x) e^{-\frac{\pi^2n^2\kappa^2}{L(\Lambda(t))^2}t},
\end{equation}
and the regularized potential is given by
\begin{equation}
V^{(reg)}(x,t)=\epsilon\tilde V(x,L(\Lambda(t))).
\end{equation}
The dynamical phase $\theta$ and the potential $\tilde V$ can be obtained from Eq.~(39) and (40) of the Ref.~\cite{MN11}. In case of the box for $\theta$ we have:
$$
\partial_x\theta=-\frac{1}{u^2}\partial_L \int\limits_{0}^{x} u^2dx.
$$
Now consider fast-forwarding of the regularized dynamics. The heat equation for such fast-forwarded system can be written as
\begin{equation}
\frac{\partial u^{(FF)}}{\partial t}=\kappa^2\frac{\partial^2 u^{(FF)}}{\partial x^2}+V_{FF}u^{(FF)},
\end{equation}
where the definition of the fast-forwarded system is given as follows \cite{MN11}:
$$
u_n^{(FF)}=u_n^{(reg)}e^{\varepsilon\theta}=u_n^{(0)}e^{\varepsilon\theta}e^{-\frac{\pi^2n^2\kappa^2}{L^2}t}. 
$$
Fast-forward ``potential'' for our case can be found in the form:
$$
V_{FF}=-\frac{d\alpha}{dt}\epsilon \theta-\alpha^2\epsilon^2\frac{\partial\theta}{\partial L}-\frac{1}{2}\alpha^2\epsilon^2(\nabla\theta)^2
$$
\begin{figure}[h!]
\includegraphics[width=16cm]{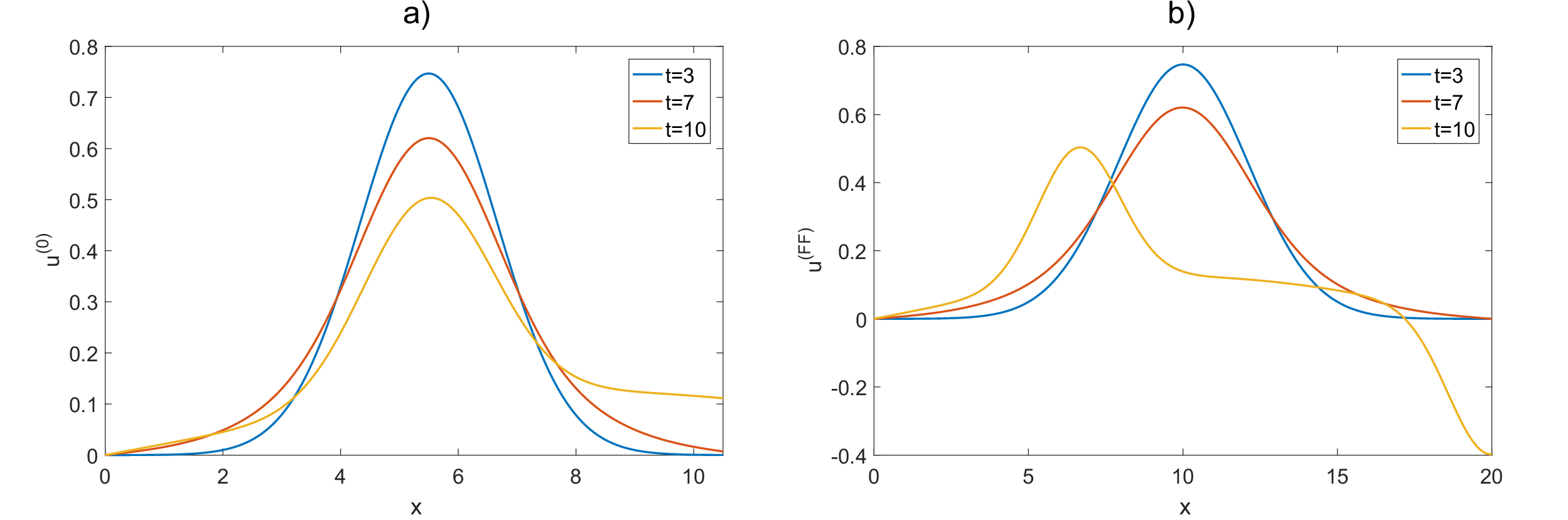}
\caption{Temperature profiles of (a) standard and (b) fast forwarded systems. The value of the heat conductance is chosen as $\kappa=0.5$, and parameter $L_0=10$ for both plots, $\varepsilon=0.04$, $\bar\alpha=100$}
\label{trp}
\end{figure}
\begin{figure}[h!]
\includegraphics[width=16cm]{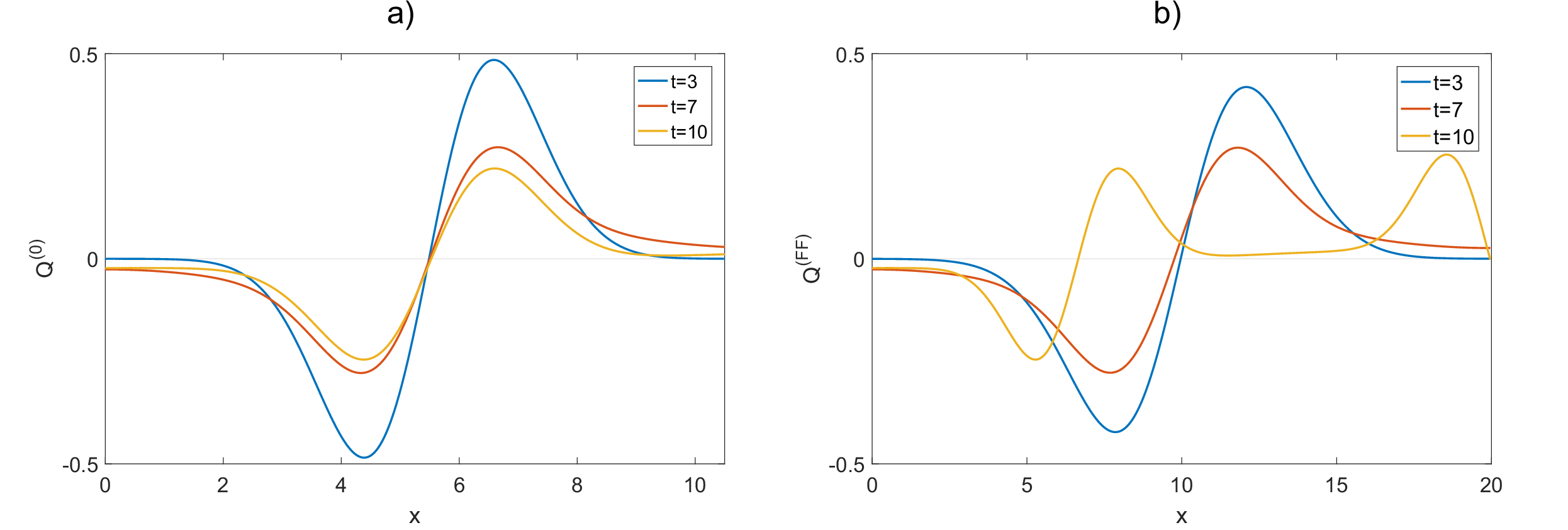}
\caption{Heat flux of of standard and (b) fast forwarded systems. The value of the heat conductance is chosen as $\kappa=0.5$, and parameter $L_0=10$ for both plots, $\varepsilon=0.04$, $\bar\alpha=100$}
\label{hflx1}
\end{figure}
\begin{figure}[h!]
\includegraphics[width=16cm]{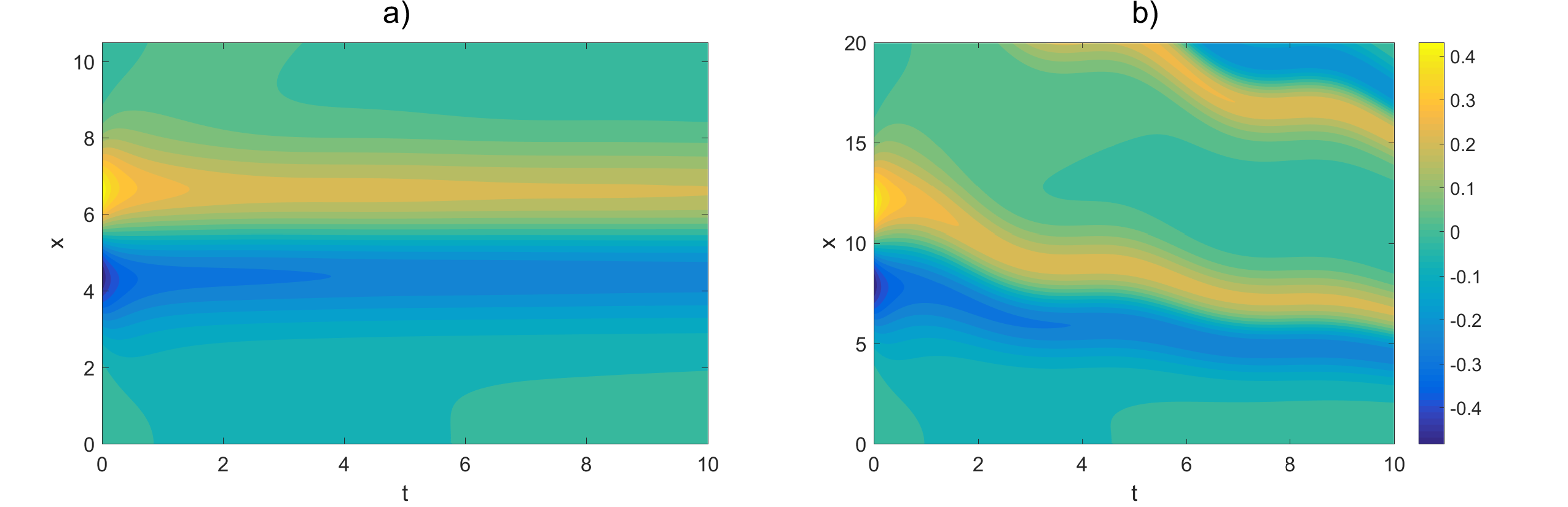}
\caption{Heat flux of of standard (a) and  fast forwarded (b) systems. The value of parameters are chosen as $\kappa=0.5$, $L_0=10$  and  $\varepsilon=0.04$, $\bar\alpha=100$ for both plots.}
\label{hflx2}
\end{figure}
\pagebreak
Figs. 1 (a) and 1 (b) present the plots of
the temperature profile for standard and fast forwarded systems, respectively. 
Abrupt qualitative difference between the two curves can be clearly seen. In particular, widening of the profile for fast forwarded state can be observed. 
Plots of the heat flux for standard and fast forwarded systems, obtained using the explicit solutions given by Eq.(29) and the numerical solution of Eq.(22), are presented in Figs. 2 (a) and 2(b), respectively. Oscillations of the heat flux becomes more intensive for fast forwarded system. Our analysis of similar plots for much longer (than that shown in the plot) time interval showed more rapid oscillations of fast forwarded state.  Figs. 3 (a) and 3 (b) present contour  plots of the heat flux for standard and fast forwarded systems vs coordinate and time for the values of parameters $\kappa=0.5$,  $L_0=10$, $\varepsilon=0.04$ and $\bar\alpha=100$. For standard system the flux is localized close to the middle of the interval, between 6 and 7, while for fast forwarded system one can observe certain blurring of the flux, although some weak localization around the points 15 and 16. Finally, one should note that in all the plots width of the box for fast forwarded case is larger than that for the standard state. This is caused by the fact that for fast forwarded system we use positive time-magnification factor that causes faster expansion of the interval in fast forwarded case. 

\section{Conclusions}
In this paper we studied  fast forward problem for heat equation by considering the latter on adiabatically expanding time-dependent box. Masuda-Nakamura fast-forward protocol \ci{mas1,mas2} is applied for  acceleration of the evolution of heat transfer process in time-dependent box with slowly moving walls. Physically important characteristics, such as temperature profile and heat flux are computed for standard and fast forwarded regimes. Considerable differences between these characteristics for standard and fast-forwarded regimes are found.  
In particular, the plots of the temperature profile for standard and fast forwarded states 
show clear quantitative and qualitative difference of the heat transfer 
 for these two regimes. Breaking of the space-periodicity in heat flux for the fast forwarded system is also found. As the physical realization of the above model one can consider  heat transport in harmonic crystals and in graphene nanoribbon subjected to time-periodic strain.  Finally, we note that
the above study can be directly extended to the cases of two- and three dimensional counterparts.

\end{document}